\documentclass[conference, 10pt]{IEEEtran}
\IEEEoverridecommandlockouts

\usepackage{amsmath,amssymb}
\usepackage{amsthm}
\usepackage{bera} 
\usepackage{bm}
\usepackage{color,soul}
\usepackage{dblfloatfix}
\usepackage[export]{adjustbox}
\usepackage{enumitem}
\usepackage{caption}
\usepackage{footmisc}
\usepackage{graphicx}
\usepackage{lipsum}
\usepackage{blindtext}
\usepackage{listings}
\usepackage{ltablex}
\usepackage{mathtools}
\usepackage{nameref}

\usepackage{tikz,multicol,wrapfig,lipsum}
\usetikzlibrary{graphs,graphs.standard}
\usepackage{pifont}
\usepackage{placeins}
\usepackage{setspace}
\usepackage{siunitx}
\usepackage{subcaption}

\usepackage{tabularx,ragged2e}
\usepackage{tikz}
\usepackage{times}  
\usepackage{textcomp}
\usepackage{url}
\usepackage{wrapfig}
\usepackage{xspace}

\newcommand{\etc}{\textit{etc.}\xspace}

\newcommand{\fig}{Fig.~}

\definecolor{cyan}{rgb}{0.0, 1.0, 1.0}

\makeatletter
\newcommand{\labitem}[2]{%
\def\@itemlabel{\textbf{#1}}
\item
\def\@currentlabel{#1}\label{#2}}
\makeatother

\colorlet{punct}{red!60!black}
\definecolor{background}{HTML}{EEEEEE}
\definecolor{delim}{RGB}{20,105,176}
\colorlet{numb}{magenta!60!black}

\definecolor{Mycolor2}{HTML}{57ADBF}
\newcommand*\circled[1]{\tikz[baseline=(char.base)]{
            \node[shape=circle,draw,inner sep=1pt,fill=Mycolor2] (char) {#1};}}

\colorlet{punct}{red!60!black}
\definecolor{background}{HTML}{EEEEEE}
\definecolor{delim}{RGB}{20,105,176}
\colorlet{numb}{magenta!60!black}

\lstdefinelanguage{json}{
    basicstyle=\normalfont\ttfamily,
    numbers=left,
    numberstyle=\scriptsize,
    stepnumber=1,
    numbersep=6pt,
    showstringspaces=false,
    breaklines=true,
    frame=lines,
    backgroundcolor=\color{background},
    literate=
     *{0}{{{\color{numb}0}}}{1}
      {1}{{{\color{numb}1}}}{1}
      {2}{{{\color{numb}2}}}{1}
      {3}{{{\color{numb}3}}}{1}
      {4}{{{\color{numb}4}}}{1}
      {5}{{{\color{numb}5}}}{1}
      {6}{{{\color{numb}6}}}{1}
      {7}{{{\color{numb}7}}}{1}
      {8}{{{\color{numb}8}}}{1}
      {9}{{{\color{numb}9}}}{1}
      {:}{{{\color{punct}{:}}}}{1}
      {,}{{{\color{punct}{,}}}}{1}
      {\{}{{{\color{delim}{\{}}}}{1}
      {\}}{{{\color{delim}{\}}}}}{1}
      {[}{{{\color{delim}{[}}}}{1}
      {]}{{{\color{delim}{]}}}}{1},
}

\definecolor{mygreen}{rgb}{0,0.6,0}
\definecolor{mygray}{rgb}{0.5,0.5,0.5}
\definecolor{mymauve}{rgb}{0.58,0,0.82}
\definecolor{terminalbgcolor}{HTML}{330033}
\definecolor{terminalrulecolor}{HTML}{000099}
\newcommand{\lstconsolestyle}{
\lstset{
	backgroundcolor=\color{terminalbgcolor},
	basicstyle=\color{white}\fontfamily{fvm}\footnotesize\selectfont,
	breakatwhitespace=false,  
	breaklines=true,
	captionpos=b,
	commentstyle=\color{mygreen},
	deletekeywords={...},
	escapeinside={\%*}{*)},
	extendedchars=true,
	frame=single,
	keepspaces=true,
	keywordstyle=\color{blue},
	morekeywords={*,...},
	numbers=none,
	numbersep=1pt,
    framerule=1pt,
	numberstyle=\color{mygray}\tiny\selectfont,
	rulecolor=\color{terminalrulecolor},
	showspaces=false,
	showstringspaces=false,
	showtabs=false,
	stepnumber=1,
	stringstyle=\color{mymauve},
	tabsize=1
}
}

\begin{document}

\title{Dissecting IoT Device Provisioning Process}

\author{\IEEEauthorblockN{Rostand A. K. Fezeu\IEEEauthorrefmark{1},
Timothy J. Salo \IEEEauthorrefmark{2}, Amy Zhang\IEEEauthorrefmark{3},
Zhi-Li Zhang\IEEEauthorrefmark{4}}
\IEEEauthorblockA{Department of Computer Science and Engineering, University of Minnesota -- Twin Cities, U.S.A. \\
\{\IEEEauthorrefmark{1}fezeu001,
\IEEEauthorrefmark{2}salox049,
\IEEEauthorrefmark{3}zhan7007,
\IEEEauthorrefmark{4}zhang089\}@umn.edu}}
\maketitle

\begin{abstract}
    We examine in detail the provisioning process used by many common, consumer-grade Internet of Things (IoT) devices.  We find that this provisioning process involves the IoT device, the vendor’s cloud-based server, and a vendor-provided mobile app.  In order to better understand this process, we develop two toolkits.  IoT-Dissect I enables us to decrypt and examine the messages exchanged between the IoT device and the vendor’s server, and between the vendor’s server and a vendor-provided mobile app.  IoT-Dissect II permits us to reverse engineer the vendor’s mobile app and observe its operation in detail.  We find several potential security issues with the provisioning process and recommend ways to mitigate these potential problems.  Further, based on these observations, we conclude that it is likely feasible to construct a vendor-agnostic IoT home gateway that will automate this largely manual provisioning process, isolate IoT devices on their own network, and perhaps open the tight association between an IoT device and the vendor’s server.  
\end{abstract}


\section{Introduction}
\label{sec:intro}

Nowadays, smart home IoT devices follow an {\em opaque}, {\em closed} "device-to-cloud" stovepipe model, i.e., {\em cloud-centric} \cite{dayalan2021veeredge, 9881629, kaala, rossposter}. A user's first interaction with a new smart home IoT device is to connect (setup) the device to their Wi-Fi for subsequent device management and control via the vendor-specific mobile app (see \fig \ref{fig:smartHome} and~\S\ref{sec:background}). This process is commonly referred to as IoT device {\em provisioning}. This provisioning process is largely a {\em manual} {\em one-device-at-a-time} task that is cumbersome, error-prone, and time-consuming. As the number of devices increases to 50 per household by 2025 \cite{g_2019_how}, this process becomes unwieldy. We are also left with lingering concerns: How trustworthy are these devices/vendors? Do they store my Wi-Fi password and other personal or private information in the cloud? How secure are these devices and vendor apps?  Might they offer an entry for hackers into my home network? 

To address these concerns, we reverse-engineer smart home IoT device provisioning process. Our goal is two-fold: i) Gain a deeper understanding of the provisioning steps to examine their security and privacy implications and ii) Explore the feasibility of developing an open, edge-centric platform for automatically provisioning smart home IoT devices. We envision that, such a platform will be co-located with a home internet gateway and provide the user with complete control in setting up policies and filtering the data flow between the home and the cloud. To this end, we concentrate only on low end consumer grade devices that are less expensive like smart light bulbs, smart switches, smart plugs \etc This is because, they are more numerous with a greater number of vendors, and are more frequently installed in homes (as opposed to devices such as smart speakers and smart video cameras).

These low end consumer grade smart home IoT devices are often provisioned using one of three modes: {\em Access Point (AP) mode}, {\em Easy (EZ) mode} or {\em Bluetooth Low Energy (BLE) mode} (See \S\ref{sec:background}). In this paper, we focus primarily on {\em Easy (EZ) mode}. This is because, 1) It is the default setup mode configured by IoT device manufacturers. and 2) It is most likely to be used by a user to manually set up his/her IoT devices. These IoT devices are able to connect to the internet because, they generally incorporate a wireless radio chipset that listens to a series of packets containing home network credentials. There are only a few wireless radio chipsets (see~\cite{li2018passwords} and \S\ref{sec:background}). All use one of these three modes to connect an IoT device to a Wi-Fi network (This makes our study applicable to a wide swath of smart home IoT devices. 

We carry out an in-depth measurement study in two stages: First, we set out to uncover the detail step-by-step provisioning process at the level of packets traversing on the network by building two test environments to bypass the vendor security mechanism employed in test devices. Second, once we have fully uncovered the detail provisioning process, we conduct a series of experiments with a goal to explore the feasibility of developing an {\em open, edge-centric} platform that will automate this process, simplify the management and control of smart home IoT devices and further empower home users with full control of their IoT devices by unlocking IoT devices from the cloud. Our key contributions are summarized below:

\textbf{$\bullet$}  We present a systematic methodology for an in-depth measurement study of smart home IoT devices. Our methodology includes, both passive packet capturing and traffic analysis as well as active reverse engineering techniques such as ``code-injection`` and ``function-tracing`` to uncover IoT provisioning process and decipher messages exchange among IoT devices, vendor clouds and mobile apps (See \S\ref{sec:howto}).
    
\textbf{$\bullet$} We design and conduct a series of experiments to explore the feasibility of breaking the "device-to-cloud" stovepipe. In particular, we develop methods for IoT device isolation and device/home network/vendor cloud segregation via creation of virtual Wi-Fi networks (with "fake" SSIDs and passwords) and an edge "proxy app" platform (See \S\ref{sec:exp}).
     
\textbf{$\bullet$} Our findings point to promising future directions for building an {\em open}, {\em edge-centric} smart home IoT framework. This envision platform will automatically manage and secure IoT devices while providing full control to smart home owners.

\section{Background and Related Work}\label{sec:background}

\begin{wrapfigure}[9]{r}{0.25\textwidth}
\centering
\vspace{-0.08in}
    \includegraphics[width=0.24\textwidth,height=0.97in]{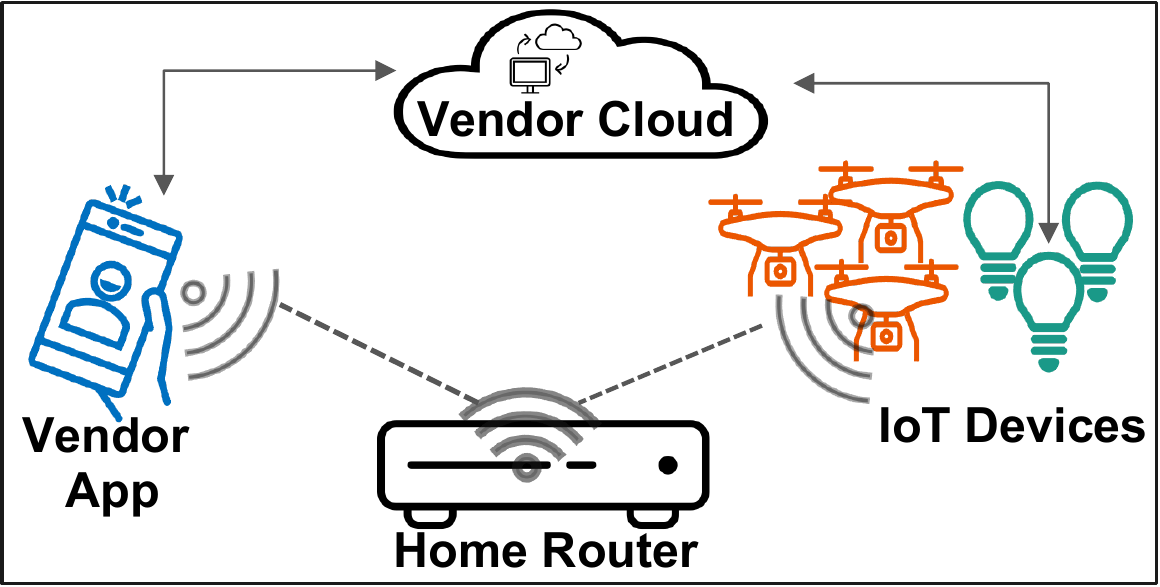}
%   \vspace*{-0.3in}
    \caption{\small Typical Smart Home IoT Device Environment.}%
    \label{fig:smartHome}
\end{wrapfigure}

Typical smart home IoT devices operate as shown in \fig \ref{fig:smartHome}. After purchasing the IoT devices, a user 
\setlength{\columnsep}{4pt}
\captionsetup{justification=centering}
must follow provisioning steps to connect the devices to their home Wi-Fi network. These steps typically include: i) The user ``must`` download to his/her mobile phone the vendor-specific app. The user ``must`` manually enter some information, perhaps about themselves. Usually, the user will be required to enter credentials for their Wi-Fi network. ii) The IoT device, the vendor mobile app, and an Internet-connected vendor cloud application communicate with each other and ready the IoT device and mobile app for operation. Generally, IoT device manufacturers use one of three modes to connect a device to a Wi-Fi during the provisioning process: \textbf{1) Access Point (AP) mode:} The IoT device appears to the mobile app to be a Wi-Fi AP. The IoT device broadcasts an SSID and the mobile app connects to the device's AP. \textbf{2) Easy (EZ) mode:} The IoT device beacons its presence.  The mobile app discovers the IoT device and sends the WLAN SSID and password.
\textbf{3) Bluetooth Low Energy (BLE) mode:}  This mode operates similar to EZ mode, except the exchanges occur over a Bluetooth network, rather than Wi-Fi network.

\subsection{Related Work}
\label{sec:related}

Several research have been conducted on smart home IoT devices and their chipsets \cite{ding2021iotsafe, ding2018safety, fernandes2016flowfence, chi2020cross}. Li, et al. identified three techniques commonly used today by IoT chipset manufacturers to send network credentials to IoT devices \cite{li2018passwords}: \textbf{1) Data in Multicast Address (DMA):} The network credential data is encoded into the last 23 bits of the destination address field of UDP packets. \textbf{2) Data in packet length (DPL):} The data is encoded in the length of UDP packets. This technique is used by the IoT devices used in this study. \textbf{3) Hybrid:} Employs both DMA and DPL to encode and send the network credentials. Other IoT research have focus on analyzing IoT traffic at scale. Most recently, Danny Yuxing, et al. \cite{huang2020iot} and M Hammad, et al. \cite{mazhar2020characterizing} conducted measurement studies on IoT devices in {\em the wild}. Their work reveals that Google and Amazon cloud platforms account for 70-90\% of the external cloud services that smart home IoT devices communicate with. 

Other papers addressed the security vulnerabilities and privacy issues of smart home IoT devices. \cite{ding2021iotsafe, ding2018safety} and \cite{dayalan2021eciot} identified new threats and security vulnerabilities that arise from the cyberspace and physical interactions of IoT devices. While they contributed to our understanding of smart home  IoT security/privacy implications and shed light on new research directions, many rely solely on "scenario analysis" and/or are based on simulations. Our work differs from and complement these existing studies by diving deep into the smart home IoT device provisioning process to understand their security/privacy implications and explore new ways for managing, controlling and securing smart home IoT devices.

\section{Uncovering Provisioning Process}
\label{sec:howto}

The first stage of our measurement study seeks to uncover the mobile app interactions with the vendor cloud, the IoT devices and the Internet. To archive this, we design two toolkits to capture and decrypt message exchanges among all three entities for in-depth analysis. First, we discuss the IoT devices under study, next we present our toolkits. 

\begin{figure}[h]
    \centering
    \vspace{-1em}
    \includegraphics[width=0.42\textwidth,height=0.8in]{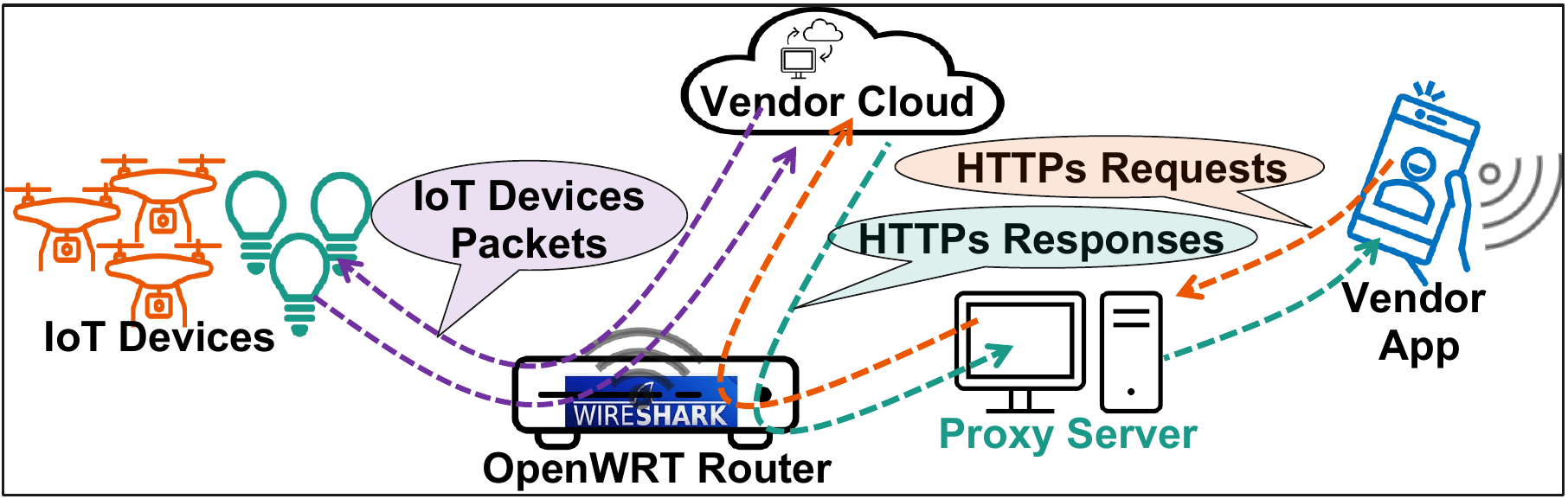}
    \vspace{-0.06in}\caption{\small IoT-Dissect I Experimental Setup}%
    \label{fig:IoT-Dissect1}
    \vspace{-2em}
\end{figure}

\subsection{Measurement Methodology and Toolkits/Testbeds}

\begin{table}[t]
    \centering
    \caption{IoT Devices Used.}
    \label{tab:devices_studied}
    \vspace{-0.5em}
    \setlength{\tabcolsep}{2pt}
    \small
    \resizebox{220pt}{!}{%
    \begin{tabular}{ || c | c | c ||}
        \hline 
        \hline
        \textbf{Chipset Models} & \textbf{Chipset Vendors} & \textbf{Device Types} \\ \hline \hline
        ESP82XX & 
        Espressif & 
        Wi-Fi LED Smart Bulbs Smart-Bulbs\\ 
        \hline
        BK7231T & 
        Beken Espressif & Dimmable Smart Bulb\\ 
        \hline
    \end{tabular}
    }
    \vspace{-2em}
\end{table}

\textbf{IoT Devices Under Study and Ethical Considerations:} In this study, we have studied multiple smart home IoT devices. In Table~\ref{tab:devices_studied} we summarize these devices' chipsets. These devices are from three different vendors, each with its own mobile app for setting up their devices. For {\em legal} and {\em security} considerations, we do not provide the specific names of the vendors. Instead in the table we have listed the type of the devices, the chipset models and vendors. 

\textbf{IoT-Dissect I:} Our testbed here, (shown in \fig \ref{fig:IoT-Dissect1}) \setlength{\columnsep}{4pt}
\captionsetup{justification=centering}
\begin{wrapfigure}[14]{r}{0.32\textwidth}
\centering
\vspace{-0.1in}
    \includegraphics[width=0.31\textwidth,height=2in]{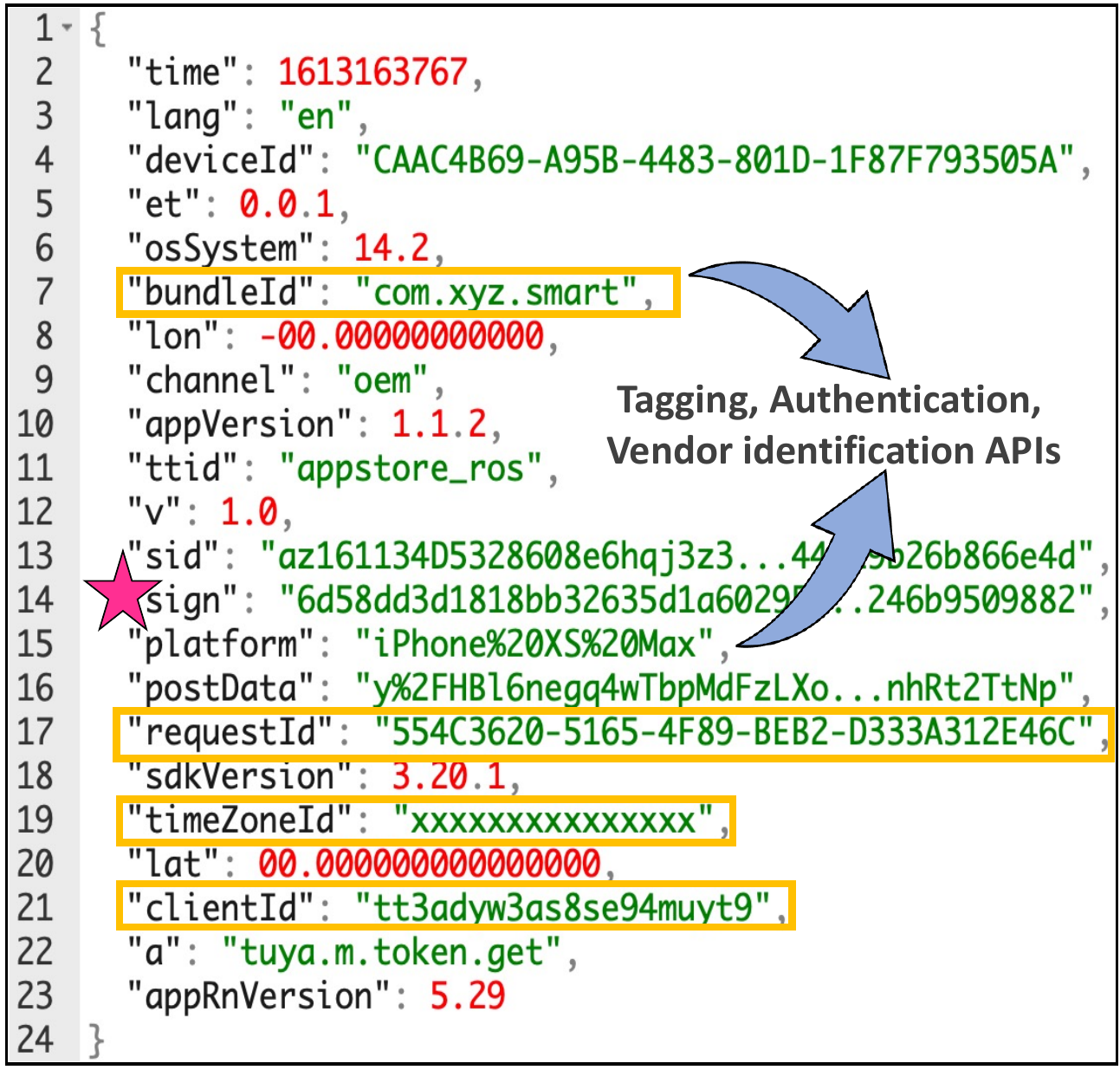}
    \vspace{-0.06in}
    \caption{\small Decrypted Post-Request Showing the Vendor Side APIs.}
    \label{fig:post_reqest}
\end{wrapfigure}consists of several IoT devices under study, a Wi-Fi network using an 
OpenWRT router (AP) instrumented with a packet-capture tool (Wireshark) and an Android phone running the vendors’ mobile apps that is also connected to the Internet. A “man-in-the-middle” (MITM) SSL proxy intercepts traffic between the Wi-Fi network and the outside vendor cloud services. As such, we are able to examine the message exchanges among the vendor cloud, mobile app and device under study.

\begin{figure*}[t]
  \centering
  \vspace{0.4em}
  \includegraphics[width=0.95\textwidth]{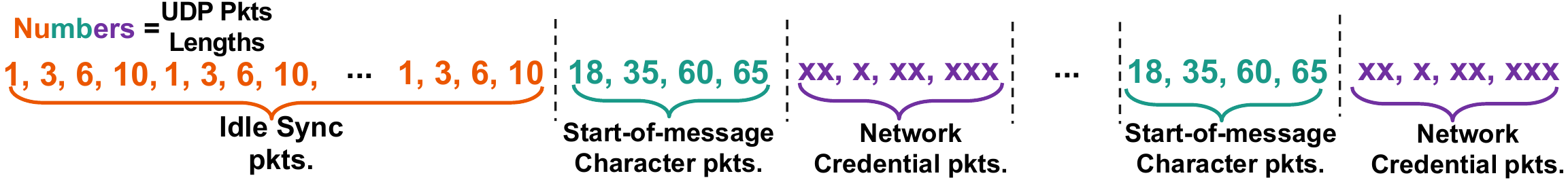}
  \caption{Broadcast UDP packets}
  \label{fig:4_udp_pkts}
  \vspace{-.25in}
\end{figure*}

\begin{figure*}[b]
    \centering
    \vspace{-1em}
\includegraphics[width=0.78\textwidth]{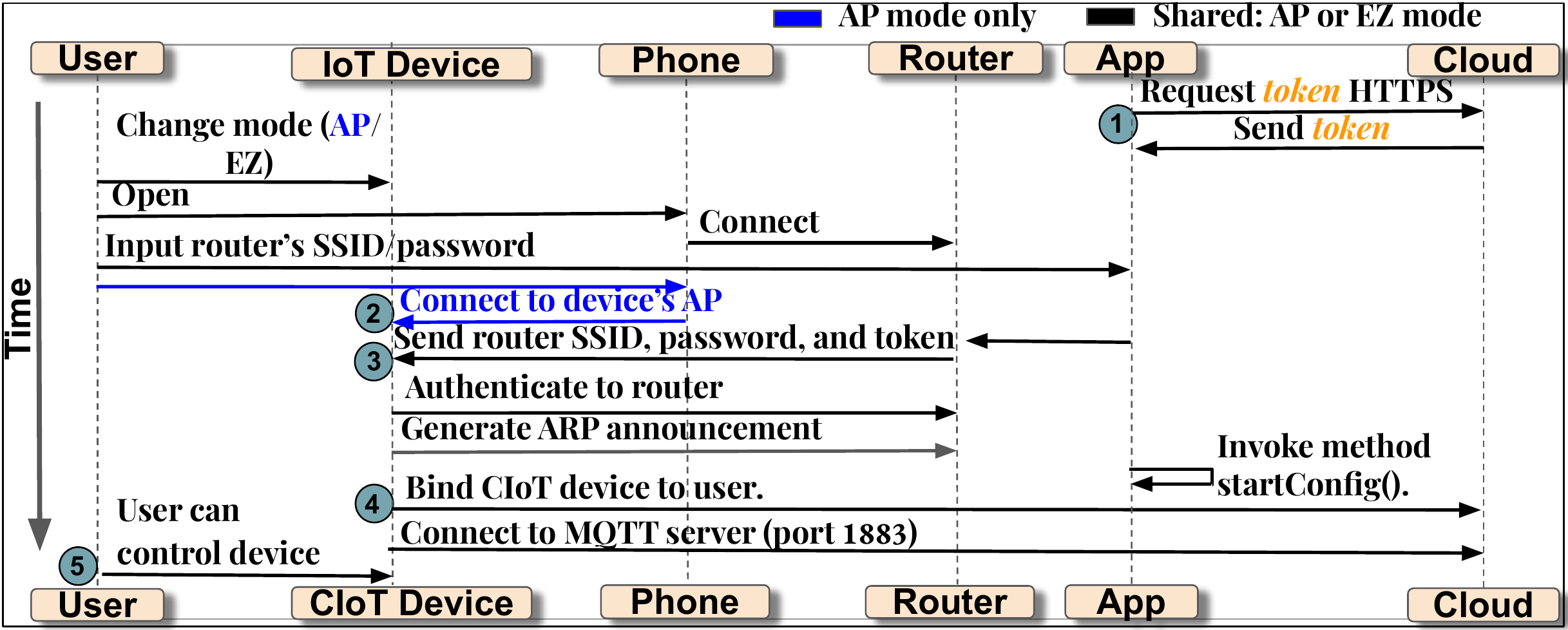}
  \caption{EZ and AP Modes IoT Device Provisioning Process.}
\label{fig:dataExchange}
\end{figure*}
\textbf{IoT-Dissect II:} For an in-depth analysis, we utilize reverse engineering techniques: Code-injection and function-tracing are used to perform  code analysis on each vendor’s mobile app (.apk files) and collect information about its operations. For instance, Function-tracing passively monitors selected function calls within the .apk files as the HTTPS requests and responses are encrypted and decrypted. Code-injection is used to extract and log the plaintext \texttt{PostData} field of each request before it is encrypted by the mobile apps and each response after it is decrypted by the mobile apps. Through this, we discover that every request and response messages between the mobile apps and vendor clouds are signed by both parties (notice the purple start \texttt{sign=} field in \fig \ref{fig:post_reqest}). This makes it extremely difficult for an MITM attacker (without access to the encryption/decryption keys) to alter the messages. In addition, all mobile apps we have studied are obfuscated to render static code analysis hard. Nonetheless it is possible to employ reverse engineering techniques to extract the security keys from mobile apps, as we have done (See \S\ref{sec:exp}).

\subsection{Preliminary Observations}
\label{observations}

Using IoT-Dissect I, we observed that the provisioning processes of all IoT devices we examined used the same initial sequence, as shown in \fig \ref{fig:dataExchange}. \circled{\textbf{1}} The mobile app first connects to the vendor cloud applications and request a provisioning \texttt{token}. \circled{\textbf{3}} The mobile app gets the user's SSID \& password and transmits that information along with the \texttt{token} encoded in a sequence of UDP packets broadcasted on port 30011 to the device. \circled{\textbf{4}} The IoT device authenticates to the user's Wi-Fi network and then binds itself to the vendor cloud services using the \texttt{\texttt{token}}. \circled{\textbf{5}} The user can now control the device using the mobile app {\em via the vendor cloud services}\footnote{In other words, the mobile app does not {\em directly} control the IoT device over the home Wi-Fi network. Hence, if a user loses Internet connectivity, the user eventually loses control of the home IoT devices, even though the home network is up and running.}.

\subsection{A Deep Dive using IoT-Dissect I}
\label{sec:dissect1}

Using the MITM SSL proxy, we uncover more details in the message exchanges between the mobile app and vendor cloud. A decrypted HTTPS post-request captured is shown in \fig \ref{fig:post_reqest}. Note that the API request parameter fields contain information about the smartphone such as the \texttt{platform}, the \texttt{clientId}, and \texttt{timeZoneId}. The vendor cloud  uses these information for tagging, authentication, validation and vendor identification purposes. The \texttt{bundleId} is used to identify the IoT device vendor;  in this example it is \texttt{com.xyz.smart} (we {\em anonymize} the vendor name here). Notice that the HTTPS post-request \texttt{PostData} API field is encrypted (we will show how they can be decrypted in~\S\ref{sec:dissect2}). We find that all vendor mobile apps under study are developed based on a third-party IoT software platform, specifically, Tuya \cite{tuya_app_sdk, smat} or Smart Life~\cite{a2017}, which is indicated via the \texttt{a} (action) field in the post requests. These third-party IoT software platforms define a set of ``standard'' APIs for managing and controlling smart home IoT devices. To confirm this finding, we  use the tool LibScout \cite{libscout} to finger-print and compare the (standard) SDK provided by the third-part IoT software platforms with those in the mobile apps. LibScout reports 
approximately 80\% or above match between the IoT software platforms' SDK and vendor mobile apps. A match of greater than 70\%  indicates that a mobile app follows  the SDK to be functional. LibScout thus verifies our findings.

\begin{figure*}[t]
  \centering
  \includegraphics[width=0.89\textwidth, height=1.3in]{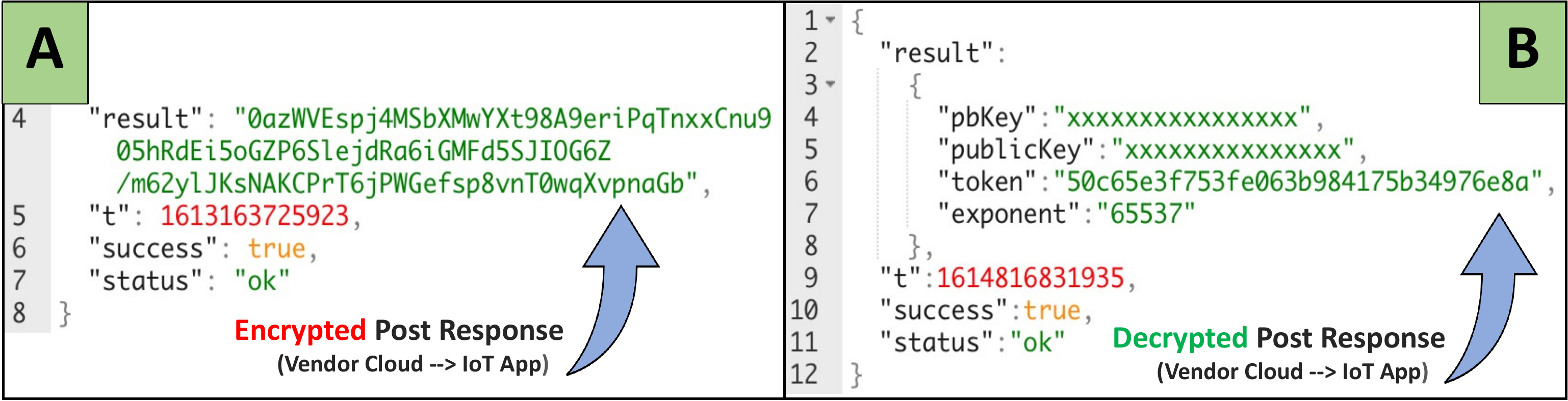}
  \caption{Post-Response from the Vendor cloud ---> IoT App. A) Encrypted \texttt{PostData} and B) Decrypted \texttt{PostData}}
  \label{fig:post_response}
  \vspace{-.2in}
\end{figure*}

\subsection{A Deep Dive using IoT-Dissect II}
\label{sec:dissect2}
Using reverse-engineering techniques, we further decrypt and dive into the message exchanges among the mobile app, IoT device and the vendor cloud. For instance, we are able to discover how the mobile apps encode the network credentials (SSID/password) and broadcast them using a sequence of UDP packets. This is shown in \fig \ref{fig:4_udp_pkts}, where this sequence of UDP packets consists of: 

\begin{itemize}[leftmargin=0.1in]
    \item \textbf{Idle sync packets:} A sequence of four UDP packets with data lengths of 1, 3, 6, and 10 bytes are broadcast repeatedly. The purpose of this packet sequence is to provide character sync.
    
 \item \textbf{Start-of-message character:} Four packets with data lengths of 18, 35, 60 and 65 bytes are broadcast. This sequence indicates the start of the data content (the network credentials). 
    \item \textbf{Network credential packets:} A series of  UDP packets with varying lengths that encode the credentials (SSID, network password, and \texttt{token}) are broadcast.
\end{itemize}

Using function tracing and code injection techniques, we are able to log the plain-text \texttt{PostData}  message before it is encrypted by the vendor mobile app and then sent to the vendor cloud as well as the decrypted responses from the cloud. Compared with \fig \ref{fig:post_response}-A where the \texttt{result} field of the post response is encrypted, \fig \ref{fig:post_response}-B shows the corresponding decrypted plain-text \texttt{PostData} response extracted after decrypted by the mobile app. We also employ function tracing technique to trace subsequent \texttt{token} functions calls. For example, we find that the \texttt{token}  generated by the vendor cloud and sent to the mobile app is subsequently encoded (using the DPL scheme) by the mobile app,  and then together with the network credentials, is passed to  the IoT device (in the form \{\texttt{SSID\_password\_token}\}) using the UDP broadcast mechanism described above.

\section{Breaking "Device-Cloud" Stovepipe}
\label{sec:exp}
Now that we have uncovered the IoT device setup process and decoded the (``secure'') message exchanges between IoT devices and the vendor cloud for provisioning, in this section we conduct a series of experiments to further understand the role of the \texttt{token} used for IoT device-cloud mutual identification and authentication, and investigate whether it is possible to break the ``device-cloud'' stovepipe. As stated in~\S\ref{sec:intro}, part of our goal is to explore the feasibility of developing an {\em open, edge-centric} platform that has the capabilities to automatically set up, manage and secure smart home IoT devices (independent of the vendor clouds), and perhaps more importantly, that endows users with full control of their own IoT devices.

\subsection{IoT Devices and \texttt{tokens}}
\label{sec:token}
In~\S\ref{sec:howto} we have seen that during the IoT device provisioning process, the message exchanges between an IoT device and the vendor cloud contain a 32-character \texttt{token} that was generated by the vendor cloud. To further understand the role of this \texttt{token} in device-cloud identification and authentication, we conduct a multi-step experiment: 1) we randomly generate a \texttt{token} of arbitrary length (x-characters), e.g., 16-character long; 2) we use either a randomly generated, 32-character \texttt{token} or an ``old'' \texttt{token} extracted from an old device-cloud message exchange; and 3) we use a recently received 32-character \texttt{token} from the vendor cloud.  We then encode the \texttt{token} produced from one of the above as well as 
the (correct) home Wi-Fi network credentials (as well as other needed information) using the DPL scheme, and generate UDP packets to {\em directly}  communicate with the IoT devices -- namely, without using the vendor mobile app. We repeat these experiments multiple times and with multiple IoT devices from different vendors. During the experiments, we run Wireshark to capture any packet generated by the IoT device and the message exchanges (if any) between the IoT device and the vendor cloud. Our findings are summarized as follows. 

In case 1), we find that during the setup process, the IoT device would inevitably indicate that it has received the Wi-Fi credentials; however, no packets are generated by the IoT device. In case 2), we find that during the setup process, the IoT device would additionally generate a message attempting to communicate with the vendor cloud. However, the device registration fails, indicating that the vendor cloud rejects the message. On the other hand, in case 3), the IoT device setup and registration process finishes successfully. Our experimental results  suggest that the IoT devices employ a rudimentary ``authentication'' process to check the validity of a vendor-generated \texttt{token} solely based on length. The \texttt{token} is primarily used by the IoT device to authenticate itself to the vendor cloud. The vendor-generated \texttt{token} is typically valid for about two hours~\cite{smart_2021_authorization}; hence any previously generated \texttt{token} older than two hours will be rejected by the vendor cloud (as a way to defend against replay attacks). We note that besides the \texttt{token}, the vendor cloud also verifies the \texttt{bundleID} (vendor ID), app metadata and the user account with the device.  It is noted in~\cite{dayalan2021eciot} that IoT devices are often locked to the vendor clouds. These experiments reveal exactly how device vendors achieve this with a \texttt{token}-based mechanism.

\subsection{Isolating IoT Devices \& Home Wi-Fi Networks} \label{sec:device-isolation}
The above experiments also illustrate a potential vulnerability of smart home IoT devices.  With the right home Wi-Fi network credentials and a recently captured \texttt{token}, a malicious user on the Internet (e.g. via breaching the user data stored in the vendor cloud)
or a compromised IoT device within a user's home, can readily take control of other IoT devices by either  hijacking the provisioning  process or the subsequent device-cloud message exchanges. Such ``\texttt{token} replay'' attacks are made possibly partially because the IoT devices always listen on open UDP port 6668 (MQTT) or UDP port 1883 (secure MQTT or MQTTs).
We therefore set out to investigate: A) whether it is possible to isolate IoT devices from other IoT devices within a home as well as isolate them from the home Wi-Fi networks; and B) whether it is possible to segment the ``direct'' device-cloud channel via a {\em trusted} intermediary (e.g., our envisaged {\em open, edge-centric} IoT management platform residing at a user's home), thus breaking the ``device-to-cloud'' stovepipe. We will describe the experiments we conduct for investigating A) below. The experiments for investigating B) are presented in~\S\ref{sec:cloud-isolation}).

\begin{figure}[b]
    \centering
    \vspace{-1em}
 \includegraphics[scale=0.4]{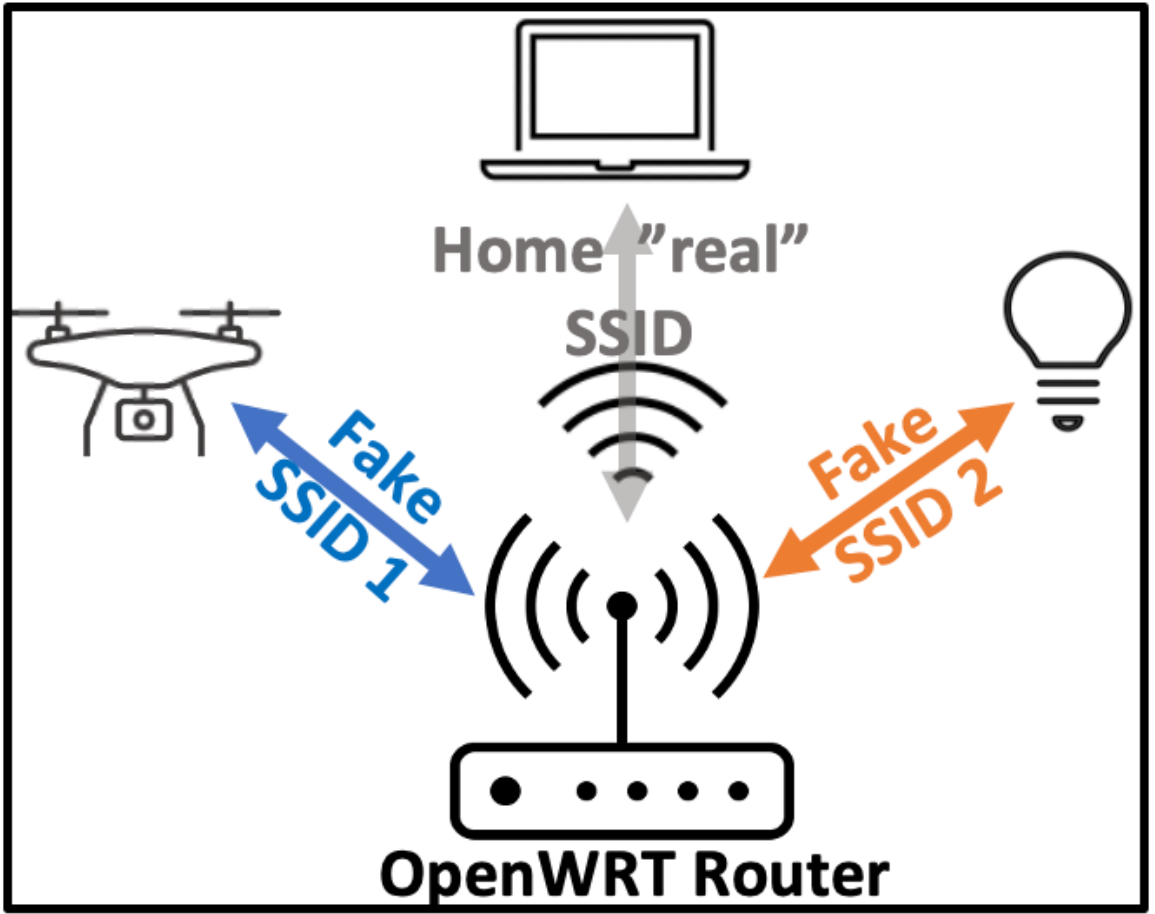}
    \caption{\small IoT device Isolation.}%
    \label{fig:fake_ssid}
\end{figure}
Our basic idea for isolating IoT devices among themselves and for separating them from home Wi-Fi networks (where other important computing and storage assets, such as desktops, laptops, reside) is to generate ``virtual'' Wi-Fi networks via ``fake'' SSIDs as shown in \fig \ref{fig:fake_ssid}.
For this we rely on programmable  Wi-Fi routers such as OpenWRT. We use an OpenWRT router to generate multiple ``fake'' SSIDs (one per IoT device or per IoT device group) and configure WPA2 encryption for each SSID. In other words, each ``fake'' SSID has its own password. We repeat a set of experiments similar to case 3) in~\S\ref{sec:token}, but this time with the ``fake'' SSID and password (instead of the true home Wi-Fi SSID and password). As in case 3) in~\S\ref{sec:token}, the provisioning process for all IoT devices  completes successfully. It is not surprising that the IoT devices have no way to distinguish the ``true'' Wi-Fi SSID used in a home Wi-Fi network from ``fake'' ones, as long as the Wi-Fi router is able to deliver messages between the device and the vendor cloud.
Hence our experimental results demonstrate that it is feasible to isolate IoT devices among themselves as well as separate them from the ``default'' home Wi-Fi networks. Clearly this relies on the ability to program a home Wi-Fi router to generate ``fake'' SSIDs or virtual Wi-Fi networks. Furthermore, this approach also enables us to prevent critical information leakage, home Wi-Fi SSIDs and passwords, as only the ``fake'' SSID and password are stored in the vendor cloud. In case of data breach, a malicious outside attacker cannot use the stolen information to break into a home Wi-Fi network to compromise, e.g., a desktop where critical user data may be stored.

\begin{figure*}[b]
\centering
\begin{minipage}{0.28\textwidth}
    \centering
    \includegraphics[width=\textwidth, height=1.6in]{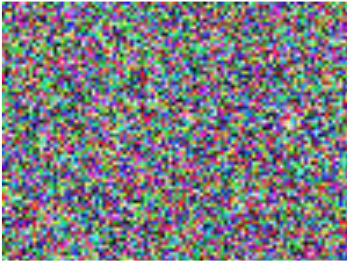}
    \caption{\small bmp file with hidden secret 2}%
    \label{fig:secret2}
\end{minipage}
\begin{minipage}{0.7\textwidth}
\centering
        \includegraphics[width=0.7\textwidth,height=1.6in]{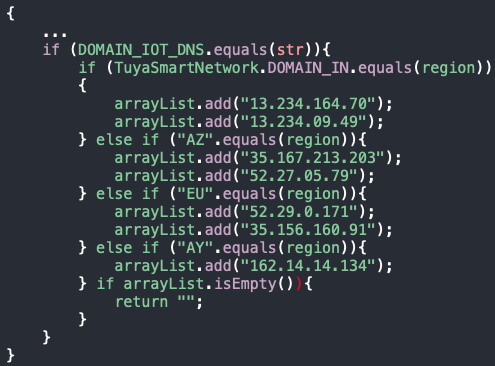}
    \caption{\small Hard-Coded DNS Server IP Addresses.}%
    \label{fig:hard_coded_dns}
\end{minipage}
\vspace*{-3ex}
\end{figure*}

\subsection{Segmenting Vendor Cloud from IoT Devices}
\label{sec:cloud-isolation}
In this last set of experiments, we develop a proxy app that runs on a local machine on the home network. It plays three roles: 1) it emulates the mobile app to interact with the vendor cloud, e.g., during the initial provision process to set up and register an IoT device; 2) it emulates an IoT device to receive commands from and exchange messages with the vendor cloud; and 3) it emulates the vendor cloud and issues control messages to the IoT device. Above, we have seen how an IoT device and the vendor cloud identify, authenticate and exchange messages with each other. The key challenge in emulating the (vendor-specific) mobile app lies in that all HTTPS post requests are signed by the mobile app using certain {\em secret keys}. We have to uncover the signing mechanism and the signing keys used by each vendor mobile app.

Just like these low end consumer grade smart home IoT devices use one of the few wireless chipsets on the market for communications (\S\ref{sec:background}), we find that the vendor mobile apps are often built on a third-party software platform (e.g., Tuya~\cite{smat} or SmartLife~\cite{a2017}) which follow a standard set of APIs for mobile app and vendor cloud interactions. We employ the same code injection and function tracing techniques used in IoT-Dissect II in~\S\ref{sec:howto} to reverse-engineer each vendor mobile app. Our experiments reveal that all the vendor apps we studied employ three secret keys for message signing: the app certificate hash string (\texttt{certHash}),  a hidden secret key (\texttt{secret2}) and an app-specific secret (\texttt{secret1}),. These keys are then concatenated as \{\texttt{certHash\_secret2\_secret1}\}, the HMAC\_SHA256 hash is computed, and the resulting hash is used to sign the post requests to the vendor cloud.  The app certificate hash string \texttt{certHash} and  \texttt{secret1} can be easily extracted using the code injection technique.

\begin{figure}[!t]
\centering
\vspace{-1em}
  \lstconsolestyle
    \begin{lstlisting}[caption={Secret 2 key obtained from bmp file. \vspace{0.1em}},captionpos=t, label={lst:secret_2}]
    > ./r_keys 8c4wxjarqdtnuju4wut5 secret2.bmp
    opening: secret2.bmp
    read 22554 bytes
    str hash: 0x7ee79371
    keys_cnt: 1 coeffs_cnt:4
     [0] offs = 0x000039ec
     [1] offs = 0x0000452a
    [KEY] [0] str: 4j8vqy4egph3thd7fdchk435hjudwsey\end{lstlisting}
    \vspace{-.1in}
\end{figure}

We found that the mobile apps, however, use steganography to hide \texttt{secret2} within a bmp file. \fig ~\ref{fig:secret2} shows a sample bmp file with the hidden secret. Nonetheless  it is possible to extract from the bmp file, \texttt{secret2} see an example in Listing \ref{lst:secret_2}. Once we uncover the message signing mechanism used by each vendor mobile app, our ``proxy'' app is able to successfully communicate with the vendor cloud, e.g., to obtain a  valid \texttt{\texttt{token}}, which can then be used to provision the IoT device for subsequent communications between the IoT device and the vendor cloud, {\em mediated via our proxy app.} We repeat the experiments with all IoT devices under study, and all are successful.

\section{Conclusions \& Future Work}
\label{sec:con}

We have conducted -- to the best of our knowledge -- a ``first-of-a-kind'' in-depth measurement study of smart home IoT devices  provisioning process. In particular, first, we have developed a systematic methodology for uncovering the mechanisms used by the IoT devices, vendor clouds and mobile apps, e.g., for identification, authentication, registration and provisioning and for analyzing the message exchanges among them. Second, we have designed experiments to understand the role of the token used in the IoT device provisioning process. Our experimental results not only reveal the potential vulnerabilities and privacy issues concerning smart home IoT devices, but also demonstrate that it is feasible to develop an {\em open, edge-centric} framework for mitigating these concerns.

\textbf{Further Discussions.} We remark that the security of IoT devices are steadily improving. More vendors are using ``secure'' protocols such as HTTPS to sign message, employing certificate validation/signing
for authentication and steganography for hiding secret keys. While we are able to reverse-engineer the
vendor mobile apps, the security contexts employed prevent simple  ``Man-in-the-Middle'' (MITM) attacks.
As first pointed out in~\cite{mazhar2020characterizing} through ``in-the-wild'' traffic analysis, many vendors also
hard-code the DNS server addresses. While the encrypted messages between IoT devices and vendor can protect user data during the transit and prevent snooping by an MITM attacker, they do not protect users from privacy leakage to the vendors, many of which collect user data and store in their clouds. As stated earlier, by breaching the vendor cloud, a malicious attacker may potentially break into a user's home network. While our proposed methods using virtual (``fake'') SSIDs/passwords for device isolation and device-network segregation can vitiate such attacks, they alone cannot prevent a more sophisticated attacker (or an ``untrusted'' vendor) to plant a malicious IoT device that can sniff wireless packets and compromise the firmware of other IoT devices to alter their behavior, due to the simple UDP broadcast encoding schemes used by the radio chipsets on these devices.

\textbf{Limitations of Our Work.}
The IoT devices under this study all use the DPL encoding scheme based on ESP8266 and BK7231T chipsets.  Our choice is partially justified by the findings in~\cite{li2018passwords} which shows that 40\% (about 329 apps/device vendors out of 821 studied) adopt this scheme. We plan to extend our study to smart home IoT devices using  DMA and other hybrid schemes.

\textbf{Future Work.} 
Our study demonstrates the feasibility of an {\em open, edge-centric}  framework for managing and securing smart IoT devices that do not rely on the vendor cloud (thus breaking the ``device-to-cloud'' stovepipe) and wrestle control from the vendor to users. This would allow users to set security and privacy policies, reduce privacy leakages and mitigate unwanted intrusions from the cyberspace. With such an {\em open} framework we envisage that ``management'' apps developed by IoT device vendors/third-parties using the framework may be automatically downloaded, validated and verified before deploying on a user's smart home platform for automatic setup and provisioning based on user-specified security/privacy policies and intent. The platform can also auto-create virtual (``fake'') SSIDs and passwords for device isolation and device/network/cloud segregation. Besides expanding our study to more diverse IoT devices, this will be a major research direction we will pursue.

\bibliographystyle{IEEEtran}
\bibliography{references}

\end{document}